\documentclass[prb,twocolumn,showpacs,preprintnumbers,amsmath,amssymb]{revtex4-1}
\usepackage{amssymb}
\usepackage{bm}
\usepackage{amsmath}
\usepackage[dvips]{graphicx}
\usepackage{color}
\usepackage{comment}
\usepackage{wrapfig}

\begin{document}

\title{Confinement of spin-orbit induced Dirac states in quantum point contacts}
\author{Tommy Li$^1$}
\affiliation{$^1$School of Physics, University of New South Wales, Sydney 2052, Australia}

\begin{abstract}
The quantum transmission problem for a particle moving in a quantum point contact in the presence of a Rashba spin-orbit interaction and applied magnetic field is solved semiclassically. A strong Rashba interaction and parallel magnetic field form emergent Dirac states at the center of the constriction, leading to the appearance of resonances which carry spin current and become bound at high magnetic fields. These states can be controlled \emph{in situ} by modulation of external electric and magnetic fields, and can be used to turn the channel into a spin pump which operates at zero bias. It is shown that this effect is currently experimentally accessible in $p$-type quantum point contacts.
\end{abstract}
%%%%%%%%%%%%%
\maketitle

One dimensional (1D) spin-orbit coupled systems have recently attracted significant interest in the context of quantum information and spintronics, playing a key role in the search for emergent Majorana fermions \cite{Majorana1, Majorana2, Majorana3} and the generation of spin polarized current \cite{SpinCurrent1, SpinCurrent2, Spincurrent3, SpinCurrent4, SpinCurrent5, SpinCurrent6}.
Interest in these systems has sparked several theoretical studies of their modified conductance properties \cite{1Ddisp1, 1Ddisp2, prec1, prec2, prec3, prec4,SanchezSerra,Xiao1,Xiao2}.
In this work I present a semiclassical solution to the scattering problem for a quantum point contact (QPC) in the presence of the Rashba spin-orbit interaction and a parallel magnetic field. The quantum states near the center of the constriction are described by a one-dimensional massive Dirac equation, with the Rashba constant and magnetic field playing the roles of the speed of light and Dirac mass respectively. The emergent fermion and antifermion states behave differently in the channel, with the latter falling into the center of the constriction, forming resonant states. This process may be considered a manifestation of Schwinger pair production, an effect which is well known in high-energy physics \cite{Schwinger, Bulanov, Ruffini}
and has recently been investigated in the context of emergent two-dimensional Dirac systems\cite{Allor, Lewkowicz}.
When the applied magnetic field is strong, the resonances become bound and generate a concentration of spin current in the channel. The properties of these states are highly sensitive to the strength of the spin-orbit interaction and the applied magnetic field, implying that they can be controlled \emph{in situ} by modulation of external fields. These results suggest an experiment in which repeated capture and release of particles inside the channel leads to a net spin polarization in the leads, transforming the channel into a spin pump and an  injector of spin current which operates in the absence of a source-drain bias. While this effect exists in principle for both $n$ and $p$ type systems, consideration of experimental parameters for existing systems suggests that hole quantum point contacts provide an ideal candidate for the realization of this effect.

A particle (either an electron or hole) moving ballistically through a QPC in the presence of a Rashba spin-orbit interaction and parallel magnetic field $B_x$ is described by scattering states of the Hamiltonian
\begin{gather} \label{hamil}
H = \frac{ p_x^2}{2m} - \alpha p_x \sigma_y - \beta \sigma_x + U(x) 
\end{gather}
where $m$ is the band mass, $\alpha$ is the Rashba coefficient, $U(x)$ is a smooth scattering potential with maximum $U_0$ at the center of the QPC ($x = 0$) and $\beta = \frac{1}{2} g \mu_B B_x$.
The spin-orbit interaction generates, in combination with the applied field, a dispersion consisting of spin-split bands $\epsilon^\pm_k$ given by
\begin{gather}
\epsilon^\pm_k = \frac{\hbar^2 k^2}{2m} \pm \sqrt{ \alpha^2 \hbar^2 k^2 + \beta^2} \ \ .
\end{gather}
For magnetic fields below a critical value $ \beta < \beta_c = m \alpha^2$, the lower band has a ``mexican hat" shape, with a local maximum at $k = 0$; the dispersion  in this case is shown in Fig. 1a. Note that the shape of the upper and lower branches near $k = 0 $ is governed by an anticrossing created by the combination of the Rashba interaction and a Zeeman interaction $0 < \beta < \beta_c$. Inside the constriction, the kinetic energy is suppressed due to the barrier potential $U(x)$. For states with certain incident energies, the scattering structure of the wavefunction is dominated by dynamics near the anticrossing in a region near the top of the barrier. In this region the linear-in-momentum term in (\ref{hamil}) becomes dominant; to leading order in $p_x$ we obtain the Dirac equation
\begin{gather}
(- \alpha p_x \sigma_y - \beta \sigma_x + U(x) ) \psi(x) = E \psi(x) \ \ .
\label{Dirac}
\end{gather}
%with $\alpha$ and $\beta$ playing the role of the \emph{effective speed of light} and \emph{Dirac} mass respectively.
We may identify the upper and lower branches of the dispersion, $\epsilon^+_k, \epsilon^-_k$ near the anticrossing with the \emph{positive energy} and \emph{negative energy} Dirac states respectively. Note that the parallel magnetic field creates a tunable Dirac gap of energy $2\beta$.

\begin{figure}
\includegraphics[width = 0.45\textwidth]{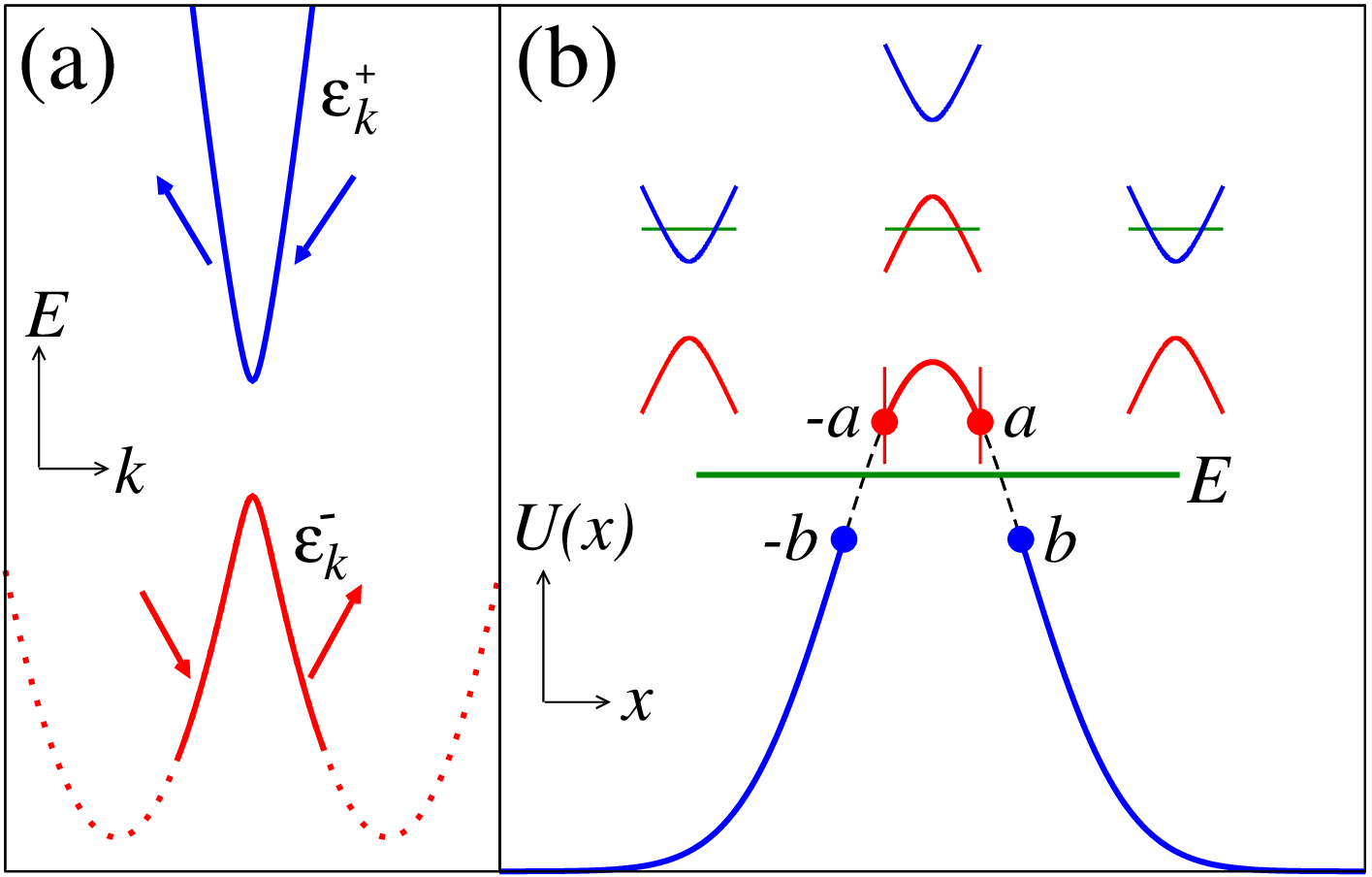}
\caption{(Color online)  Confinement of a Dirac fermion in a QPC, which occurs when the parallel magnetic field is below the critical value $\beta = \frac{1}{2} g \mu_B B_x < \beta_c$. a) The 1D dispersion near the anticrossing, with the bold lines showing the region in which Dirac fermion behaviour is expected to appear. The two spin branches $\epsilon^+_k, \epsilon^-_k$ are identified as the positive and negative energy Dirac states respectively. Arrows indicate the direction of spin polarization. b)  The potential barrier $U(x)$ (lower curve), and the situation in momentum space (three upper figures). There are three regions which form the scattering wavefunction. In the asymptotic region $|x| > b$, the kinetic energy lies in the upper (positive energy) branch. Inside the barrier, $|x| < a$, the kinetic energy lies in the lower (negative energy) branch. Quantization of motion in the region bounded by vertical lines gives rise to a subbarrier resonance. The energy is indicated by the horizontal line. The tunnelling regions are shown with dashed lines. }
\end{figure}

Transmission through the constriction differs qualitatively in the situations $\beta > \beta_c$ and $\beta < \beta_c$. In the former, the transmission probability in each band simply drops to zero as the Fermi energy successively crosses the bottom of each band, leading to the well-known double-step in the channel conductance which is the experimental signature of Zeeman split subbands \cite{Patel}. The second situation, $\beta < \beta_c$, is highly non-trivial due to the possibility of tunnelling from a positive to a negative energy state when the barrier is sufficiently high, $U_0 > E + \beta$. Tunnelling occurs at the sides of the potential barrier (indicated by dashed lines in Fig. 1b) in the regions where the kinetic energy lies within the Dirac gap.  The result is the appearance of a particle in a negative energy state at the top of the barrier, which is characterized by a negative effective mass, $\frac{ d^2\epsilon_k}{dk^2} < 0$ (or equivalently, positive effective mass but opposite electric charge) and therefore sees the repulsive potential as a quantum well. This demonstrates that there exist two types of charge carriers in the channel, which exhibit qualitatively different behaviour in the electrostatic potential $U(x)$: the positive energy states in the branch $\epsilon^+_k$  fall away from the barrier, while the states in the \emph{antifermion} branch $\epsilon^-_k$ fall towards the barrier. This process is illustrated in Fig. 1b. The lower curve shows the potential $U(x)$, with the tunnelling region shown with dashed lines. The scattering wavefunction is formed by two pairs of turning points $x = \pm b$ (assuming a symmetric barrier) and $x = \pm a$. The situation in momentum space is shown in the three figures in the upper part of the same panel. In the regions $|x| > b$, the kinetic energy (shown with a horizontal line) lies in the upper branch, and the particle accelerates away from the barrier. In the region $|x|< a$, the kinetic energy lies in the lower branch. In this region the particle behaves as an antifermion; it falls into the barrier becomes confined. Quantization of motion between the turning points $x = \pm a$ gives rise to a resonant state; the energy of the resonant state is shown by the horizontal line intersecting the lower curve in Fig. 1b.  Note that the kinetic energy of the ``antifermion" must remain above the bottom of the lower branch $E_{min}$ in order for free motion to persist in the confinement region; the particle is expected to behave as a Dirac fermion when $E - U(x) > E_{min}$. The range of energies in which resonances are expected to exist is given by
\begin{gather}
E+ | E_{min}| > U_0 > E + \beta \ \ , \ \ E_{min} =  - \frac{ m \alpha^2}{2} - \frac{ \beta^2}{2m \alpha^2} \ \ .
\label{condition}
\end{gather}

The existence of two species of carriers which accelerate in opposite directions in the same electric field may be regarded as a manifestation of Schwinger's mechanism for the spontaneous generation of fermion-antifermion pairs in strong electric fields\cite{Schwinger}. While ubiquitous in quantum electrodynamics\cite{Bulanov,Ruffini}, direct observation of pair production requires electric fields $e E > \frac{m^2 c^3}{\hbar} \gtrsim 10^{16} \text{V cm}^{-1}$ which to date have only been accessible in high-energy collisions\cite{SLAC}.
%\color{red}nuclear decay, photon-photon scattering and so on\cite{nonlinearity}.\color{black}
In the condensed matter context, this effect has been studied theoretically in graphene \cite{Allor,Lewkowicz} and is closely related to the phenomenon of atomic collapse recently observed in the material \cite{AtomicCollapse}. In our case, the Schwinger mechanism distinguishes the localized ``Dirac" state from quasibound states arising from double barriers, edge or multichannel effects as previously studied \cite{SanchezSerra,Xiao1,Xiao2,prec3}, and as we shall see later, gives them properties which will enable them to facilitate pumping of spin across the channel.

The transmission probability for a state incident in the upper branch $\epsilon^+_k$ may be derived by consideration of the explicit structure of the scattering wavefunction.
The wavefunction consists of an asymptotic wave with positive energy which undergoes reflection at the sides of the barrier, $x = \pm b$, and a negative energy part corresponding to free motion inside the barrier, $|x| < a$. The positive energy and negative energy components of the wavefunction are coupled by tunnelling in the regions $a < |x| < b$, leading to a finite lifetime which is proportional to the inverse tunnelling rate. Approximating the barrier by a linear function, $U'(x) = \lambda$, the resonant width is given by $\tau^{-1} \propto e^{-2\pi \gamma}$ where the exponent $\gamma$ is given by Schwinger's formula\cite{Schwinger} upon identification of $\beta$ with the rest energy  and $\alpha$ with the effective speed of light:
 \begin{gather}
 \gamma = \frac{ \beta^2}{ 2\alpha \lambda \hbar} \ \ .
 \label{gamma} 
 \end{gather}
Accounting for periodic motion in the confinement region $|x| < a$ and tunnelling, we obtain explicitly for the transmission probability
\begin{gather}
T = |\frac{ 1}{ e^{ 2\pi \gamma} + (e^{ 2\pi \gamma} - 1) e^{ i \oint{ kdx} -i \delta \varphi} } |^2 \ \ ,
\label{trans}
\end{gather}
where $\oint{ kdx}$ is the semiclassical phase acquired by the ``antifermion" over one period of semiclassical motion. In the tunnelling regions, the solution of the Dirac equation (\ref{Dirac}) may be expressed analytically in terms of the parabolic cylinder functions, which leads to a phase shift  $\delta \varphi = 2\text{Arg} \Gamma(i \gamma) - 4 \gamma \left[ \ln 2 \sqrt{\gamma} - 1 \right] + \frac{\pi}{2}$. This phase shift appears in the WKB quantization condition for the existence of a standing wave, $\oint{ kdx} = 2\pi(n + \frac{1}{2} + \delta \varphi)$ at energies corresponding to the location of the Breit-Wigner resonances.
We may regard (\ref{trans}) as the general expression for the transmission probability for a one-dimensional Dirac fermion in a smooth potential barrier. 
In the limit of strong magnetic fields $\beta \gg \sqrt{\alpha \hbar \lambda}$, the resonant states become bound. 
In the opposite limit $\gamma \rightarrow 0$, resonant tunnelling occurs over a broad range of energies, reflecting the Klein paradox for ultrarelativistic Dirac fermions \cite{Klein,KleinGraphene}. Note that we have considered a state with asymptotic energy in the upper branch, $\epsilon^+_k$; in the semiclassical approximation a particle incident from the lower branch $\epsilon^-_k$ does not undergo reflection at the Dirac gap and the corresponding transmission probability is non-resonant.

Approximating the QPC potential as a parabolic barrier, $U(x) \approx U_0 - \frac{ m \omega^2 x^2}{2}$, the resonant spectrum is equivalent (\emph{upon reversing the sign of the energy}) to that of a harmonic oscillator with mass $|m^*|$ and oscillator frequency $\omega^* = \sqrt{| \frac{m}{m^*}|}\omega$,
\begin{gather} 
E_n = U_0 - \beta - \omega^* (n + \frac{1}{2})  \ \ .
% \ \ , \nonumber \\
% \ \ \ \omega^* = \sqrt{-\frac{m}{m^*}} \ \omega \ \ .
\label{spectrum}
\end{gather}
The $n = 0$ mode possesses the highest energy, with higher modes forming an inverted tower of oscillator states extending downward in energy. In deriving the spectrum (\ref{spectrum}) we assume that the lower band may be approximated by a quadratic dispersion $\epsilon^-_k \approx - \beta + \frac{ \hbar^2k^2}{2m^*}$, so that in the semiclassical picture the resonant states correspond to simple harmonic motion confined between two turning points $x = \pm a$. (In this limit we also have $\delta \varphi \rightarrow 0$.) The spectrum terminates at finite $n$ due to the condition (\ref{condition}). This condition may be expressed in terms of parameters $\zeta = \frac{ \omega}{ m \alpha^2}, \ \ \eta = \frac{ \beta}{m \alpha^2}$ as
\begin{gather}
\zeta < \frac{1}{2n + 1} (1 - \eta)^\frac{3}{2} \eta^\frac{1}{2}\ \ .
\label{number}
\end{gather}
The number of quasibound states predicted by (\ref{number}) in different regions of the space of parameters ($\eta, \zeta$) is shown in Fig. 2. In order to observe resonances we require the variation of the potential inside the channel to be smooth compared to the energy scale of the spin-orbit interaction, $\zeta \ll 1$. The optimal regime, in which a large number of particles is trapped in the constriction, occurs when $\eta \approx \frac{1}{4}$. Note that the number of quasibound states is much less sensitive to the size of the magnetic field than to the shape of the confining potential.

It should be noted that number of states given by (\ref{number}) may undergo modification  in the presence of electron-electron interactions. In the Hartree approximation, the presence of a localized state at the center of the QPC both renormalizes the barrier height $U_0$ and makes the barrier sharper near the top (increasing $\omega$), which would na\"{i}vely be expected to reduce the number of resonances. At the same time, however, the spin-orbit coupling may be enhanced significantly via the exchange interaction in a similar manner to the exchange enhancement of the $g$-factor in quantum wires\cite{Wang,Martin}. Thus a detailed calculation beyond the Hartree approximation is required to account for the modification to (\ref{number}) by interactions.

\begin{figure}
\includegraphics[width = 0.45\textwidth]{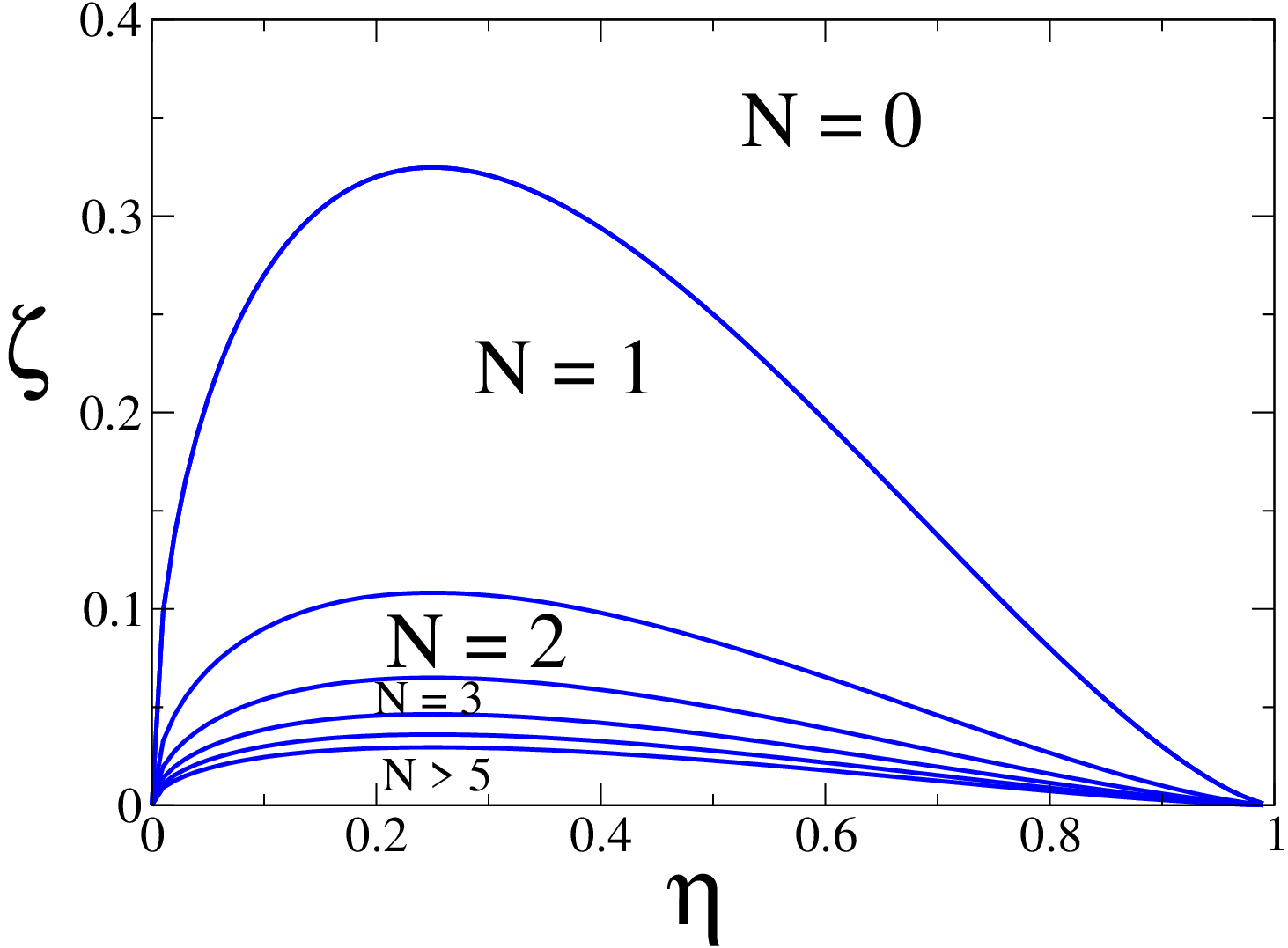}
\caption{(Color online) Predicted number of quasibound states for parameters $\eta, \zeta$, given by Eq. (\ref{number}).}
\end{figure}

Let us now address the possibility of observing resonant Dirac states in the typical experimental situation. Near pinchoff, the barrier height is tuned to the Fermi energy, $U_0 \approx E_F$. We may parametrize the barrier in terms of the QPC length, $ U_0 = \frac{1}{2} m \omega^2 ( \frac{ l}{2})^2$, to obtain
\begin{gather}
\zeta = 
%\frac{\hbar}{m \alpha^2} \sqrt{ \frac{ 8 E_F}{ m l^2}}=
 \frac{ 2}{ k_F l}  (\frac{ \hbar k_F}{m \alpha})^2 \ \  .
\end{gather}
The value of $l$ is limited by the ballistic mean free path, which is approximately one micron in 2D GaAs. Taking $ l = 1 \mu \text{m}$, at typical experimental density $n = 10^{11} \text{cm}^{-2}$ in the 2D regions, we have $k_F l \approx 10^2$; in the strongly spin-orbit coupled materials InSb and InAs, the conduction band effective mass ($m \approx 10^{-2} m_e$) is prohibitively small despite the strong Rashba effect, since $ \frac{ \hbar k_F}{m \alpha} \approx 10^{2}$ at typical electric fields $\alpha \propto E_z = 10^6 \text{V} \text{m}^{-1}$. While in principle the 1D band-mass may be renormalized by inter-subband coupling, the initial value $\zeta \sim 10^2$ would require a renormalization by more than two orders of magnitude. %, making observation of resonant Dirac fermion states difficult.
%While this effect is parametrically unobservable in electron systems in absence of a large renormalization of the 1D band-mass,

Alternatively, we may consider $p$-type systems which possess a much larger value of $m$ and a significantly stronger spin-orbit interaction\cite{WinklerBook}.
\begin{comment}
In the hole case the Rashba interaction in the Hamiltonian (\ref{hamil}) originates from a 2D Hamiltonian which is cubic rather than linear in momentum \cite{Winkler2002}
\begin{gather} \label{cubic}
 H^{(2D)}_R = \frac{ i C}{2} ( \hat{p}_+^3 \sigma_- - \hat{p}_-^3 \sigma_+)
\end{gather}
where $\hat{p}_\pm = \hat{p}_x \pm i \hat{p}_y$, and $C$ is a coefficient parametrizing the asymmetry of the well ($C = 0$ for a symmetric well). The corresponding 1D Hamiltonian may be obtained after imposing lateral confinement; we may make the replacement
 $\langle p_y \rangle = 0, \ \  \ \langle p_y^2 \rangle \approx  (\hbar k_F)^2$, giving
\begin{gather} 
H^{(1D)}_R = -3C p_F^2 \hat{p}_x \sigma_y + C \hat{p}_x^3 \sigma_y \  \ \ .
\end{gather}
The first term is the Rashba interaction, $\alpha \hat{p}_x \sigma_y$ with $\alpha = 3 C p_F^2$ while the second term provides a qualitatively unimportant modification of the dispersion.
\end{comment}
In $p$-type point contacts, the size of the 1D Rashba coefficient may be related to the Rashba coefficient in the 2D hole gas. The latter may be determined from the zero-field spin splitting observed in magnetic oscillations; we obtain values $0.6 \leq \frac{m\alpha}{\hbar k_F} \leq 1$ from such studies in GaAs-AlGaAs heterojunctions \cite{Grbic, Nichele}. Taking typical values\cite{effectivemass} for the hole density $n = 10^{11} \text{cm}^{-2}$ and $m = 0.4 m_e$, we obtain $\eta = \frac{1}{5}, \zeta = \frac{1}{18}$ at $g B_x = 3 \text{T}$ for the lower value $m \alpha = 0.6 \hbar k_F$. In this regime (\ref{number}) predicts $N = 3$.

For comparison with the semiclassical solution described earlier, the scattering problem was also solved by explicit numerical integration of the Schrodinger equation, with the corresponding QPC conductance shown in Fig. 3. The potential barrier was modelled as a Gaussian, $U(x) = U_0 e^{ - \frac{ x^2}{w^2}}$ with $w =  0.5 \mu\text{m}$, and the conductance is shown for values of $g B_x = 1\text{T}, 2\text{T}, 3\text{T}$ (note that the typical $g$-factor in hole quantum point contacts is $g \approx 0.5$\cite{Chen}). While the transmission problem was solved with the complete Hamiltonian (\ref{hamil}), the numerical solution clearly exhibits the signatures of Dirac physics which were described earlier in semiclassical language based on the approximation (\ref{Dirac}). Results are shown in the case when the Fermi energy is high enough to intersect both branches of the dispersion in the asymptotic region; the conductance is given by the sum of transmission probabilities for particles injected in the upper and lower branches.

For a particle injected in the lower branch $\epsilon^-_k$, the transmission probability smoothly decreases to zero when $E - U_0 < E_{min}$, at which point there is reflection from the bottom of the lower band; the lower branch therefore contributes a conductance $G_- = \frac{e^2}{h}$ in the range (\ref{condition}). The upper branch contributes a conductance $G_+ = \frac{e^2}{h} T$ where the resonant transmission probability $T$ is given by (\ref{trans}).
%The resonances appear on the plateau $E_F - \beta < U_0 < E_F + |E_{min}|$, in accordance with the condition (\ref{condition}).
Three resonances are shown as predicted by Eq. (\ref{number}), with the roughly equal energy spacing reflecting the oscillator spectrum (\ref{spectrum}). The lifetime of the rightmost resonance on the solid trace in Fig. 3 (indicated by the arrow) was calculated to be $\tau = 7.4 \times 10^{-10} \text{s}$. The lifetime, according to (\ref{trans}) and (\ref{gamma}), displays an exponential dependence on both $\alpha$ and $\beta$, which may be used to provide a highly sensitive measurement of the Rashba coefficient. Upon increasing the magnetic field from 1T to 3T, the resonances become significantly narrower (Fig. 3). At 5T, the lifetime is $\sim 10^{-8} \text{s}$, and at 10T, the lifetime is $\sim 10^{-1} \text{s}$. In this limit an additional broadening of the resonance is expected at finite temperature due to inelastic electron-electron collisions in the channel, nevertheless our analysis shows that at high magnetic fields, the resonance is effectively bound. Note that while varying the magnetic field from 1T to 3T significantly narrows the resonant peaks, it does not change the number of resonances. This distinguishing behaviour, which was derived earlier (see Eq. (\ref{number})) from the inverted harmonic oscillator spectrum (\ref{spectrum}), permits a straightforward identification of the effect in experiment.

\begin{figure}[t]
	\includegraphics[width = 0.4\textwidth]{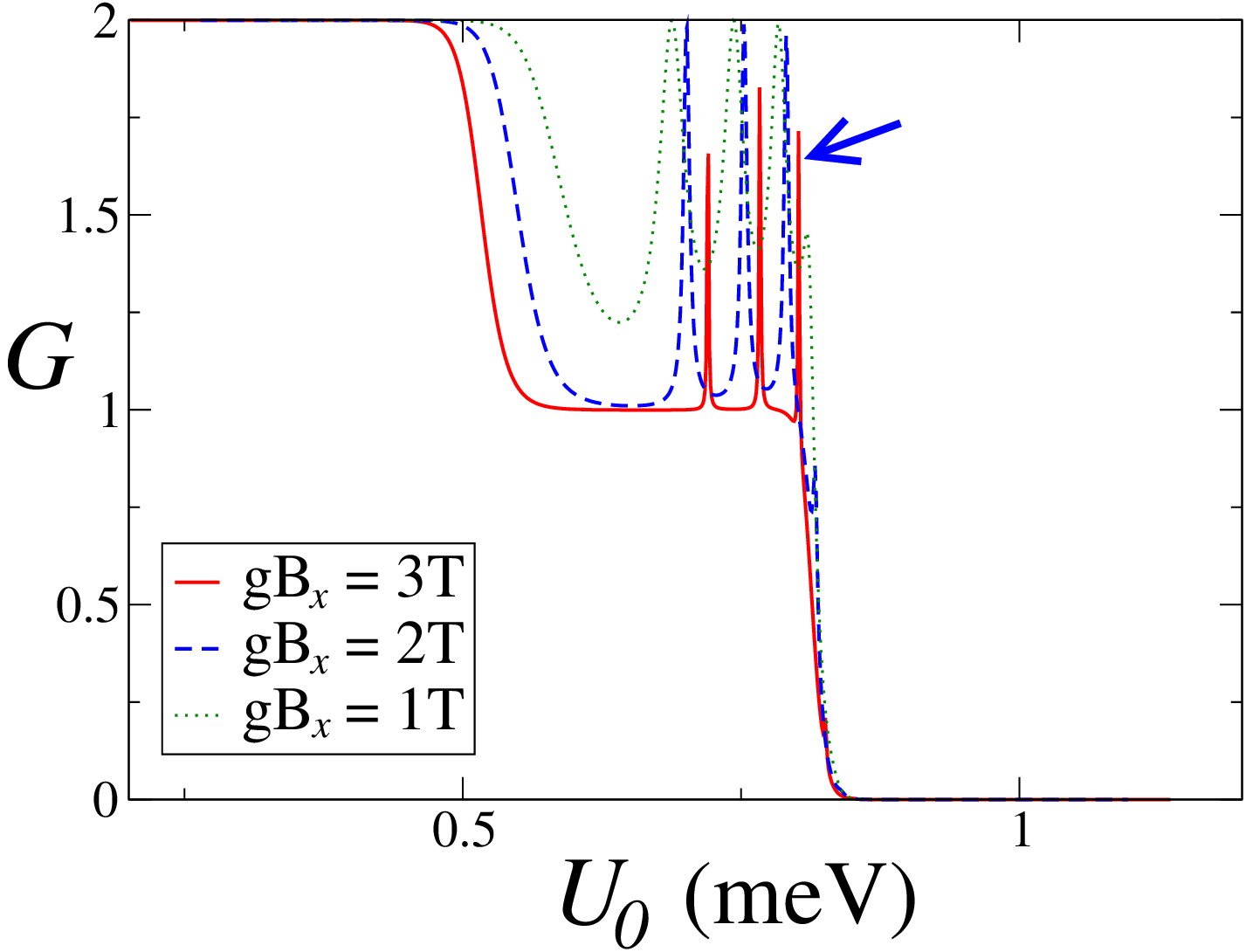} 
	\caption{(Color online) The QPC conductance in units of $\frac{e^2}{h}$ for a gaussian barrier $U(x) = U_0 e^{ - \frac{ x^2}{w^2}}$ with parameters $w = 50  \text{nm}$, $E_F = 0.6 \text{meV}$, $\frac{m\alpha}{\hbar k_F} = 0.6$, and values of magnetic field $g B_x = 3\text{T}$ (solid, red), 2T (dashed, blue), 1T (dotted, green). The spin-resolved current density corresponding to the peak in the solid trace indicated by the arrow is shown in Fig. 4.}
\end{figure}

Let us finally consider the spin structure of the bound state. At low values of $\eta$ (i.e. $\beta \ll m\alpha^2$), the wavefunction inside the barrier is proportional to $\psi(x) \propto e^{i \int{ k dx}} |+ \rangle_y + e^{-i \int{ k dx}} |-\rangle_y$ where $|\pm \rangle_y$ are spinors with polarization along the $\pm y$ axis. In such a state, both the current and magnetization are zero. Instead of a total spin, the state carries a total spin current $J_{x \mu} (x) = \frac{1}{2} \{ v_x, s_\mu\}$ which is concentrated at the top of the barrier.
The existence of a localized region of spin current is a consequence of the fact that, in the presence of a Zeeman gap, only states of one chirality participate in the tunnelling process which leads to localization. In this sense the bound states considered here bear a strong similarity to chiral subgap (edge and surface) states in topological insulators\cite{edge}, although they are supported by a smoothly varying potential rather than an edge. The current densities of the spin-up and spin-down components of the wavefunction are shown in Fig. 4a for the wavefunction at the resonance indicated by the arrow in Fig. 3.
\\ ~ \\

\begin{figure}[h]
\includegraphics[width = 0.45\textwidth]{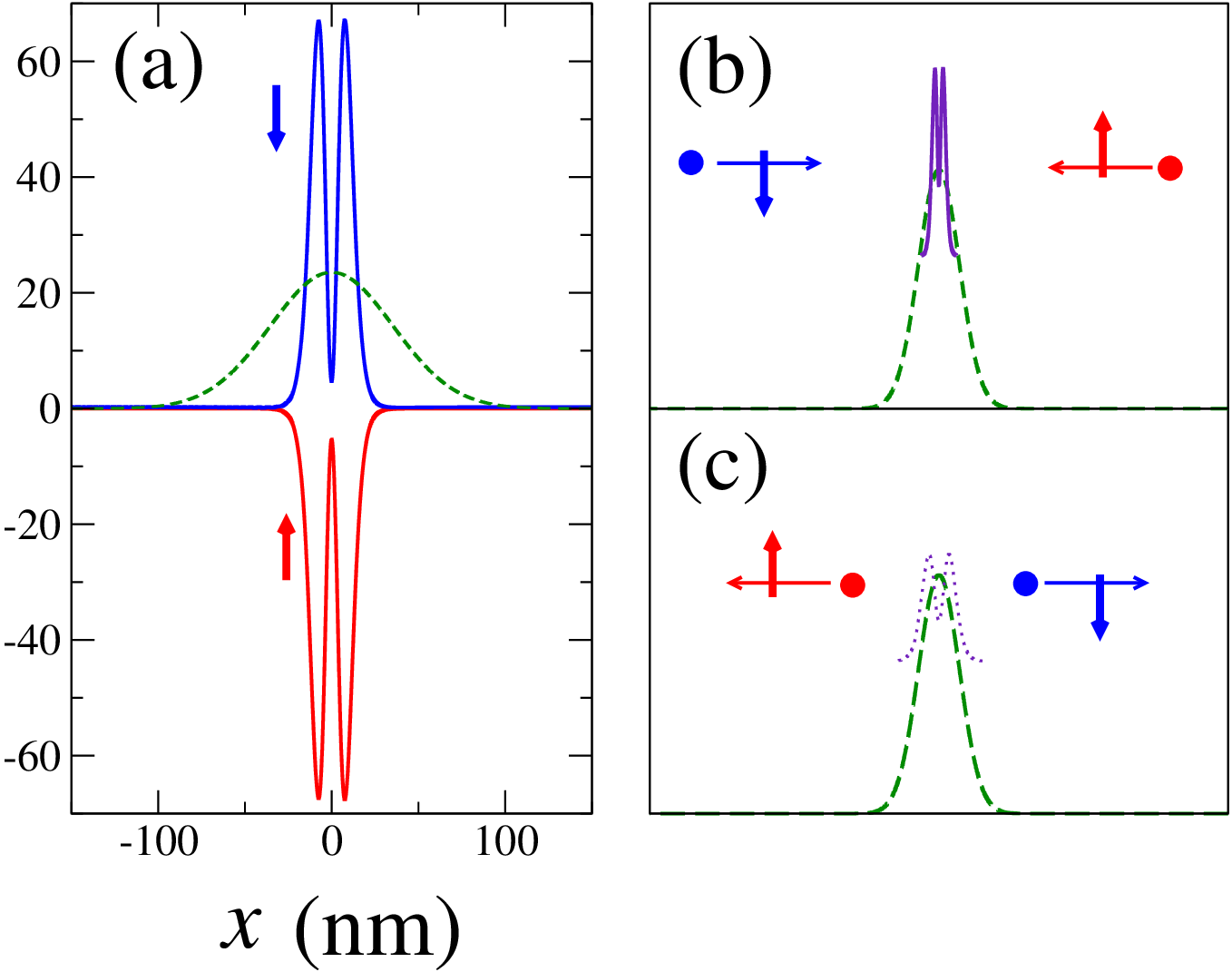} 
\caption{(Color online)
	% The probability and spin current densities for the wavefunction corresponding to the resonance indicated in Fig. 3. The only nonvanishing component of the spin current is $J_{xy}$.
	Spin current and spin pumping via the bound state. a) The current densities (arb. units) in the spin-up and spin-down components of the wavefunction at resonance, $J_{x+} = \psi^*_+ J_x \psi_+$ (red, positive curve) and $J_{x-} = \psi^*_- J_x \psi_-$ (blue, negative curve), where $\psi = \psi_+ |+\rangle_y + \psi_- |-\rangle_y$. The barrier is shown by the dashed line. Since the state is localized, the total current everywhere is zero, however the state possesses a spin current $J_{xy} = \frac{J_{x+} - J_{x-}}{2}$. b) Upon adiabatically switching on the magnetic field, a particle is captured inside the barrier from the left lead, initially in a spin-down state. c) When the magnetic field is switched off instantaneously, the resonance decays, either returning the particle to the left lead in a spin-up state, or transferring it to the right lead in a spin-down state. The mirror process occurs for a particle injected from the right lead. This process increases the number of spin-up carriers in the left lead and the  number of spin-down carriers in the right lead, i.e. it pumps spin across the channel in the absence of DC current.
	 }
\end{figure}

It is straightforward to show that the localized spin current in our situation may be used to pump spin across the channel. Let us consider the spin states in the positive and negative energy bands $\epsilon^\pm_k$ near the anticrossing. In the positive energy branch, states with positive momenta have spin tilted towards the $-y$ axis, while states with negative momenta have spin tilted towards the $+y$ axis. In the negative energy branch, the situation is opposite (see Fig. 1a).
Let us consider the component of spin current $J_{xy}$ which is given by
\begin{gather}
J_{xy} =\frac{1}{4} \{ \frac{p_x}{m} - \alpha \sigma_y, \sigma_y \} = \frac{p_x}{2m} \sigma_y -\frac{\alpha}{2} \ \ .
\label{spincurrent}
\end{gather}
When $\alpha > 0$ the first term is negative for a positive energy state and positive for a negative energy state. Thus during the capture of a particle from the leads, the spin current of the particle is increased.
In general, the bound state may be controlled by modulation of the parameters $\alpha, \beta$ which implies that the QPC can function as a spin transistor.
We may perform, for example, a gedaken experiment in which the external magnetic field is first switched on slowly, resulting in the capture of a particle into the channel (assuming that the system is tuned so that the Fermi energy coincides with a resonance). During this process a positive energy state in the reservoirs is converted into a negative energy state in the constriction, and the spin current of the system is increased. Let us now instantaneously switch off the external magnetic field. Since the external field is required for trapping, the bound state decays and the particle is transferred into either the left or right reservoir. However, in the absence of the external magnetic field, the operator of spin current (\ref{spincurrent}) commutes with the Hamiltonian (\ref{hamil}). Thus the initial spin current generated by adiabatically switching on the magnetic field persists even in the absence of a localized state. It is straightforward to show that this process either transfers a particle from one reservoir to the other without changing its spin, or returns a particle to its original reservoir with a spin flip, as shown in Fig. 4b,c. Cycling the applied field in this manner pumps spin from one reservoir to the other, generating a non-equilibrium polarization in the leads.

These results demonstrate that, aside from interactions and multichannel effects, the single particle, single mode transmission problem is in itself highly nontrivial in 1D spin-orbit coupled systems. For the confinement mechanism described, only a strong spin orbit interaction, a smoothly varying repulsive potential, and an external magnetic field are required. Furthermore, these results illustrate a surprising manifestation of Dirac physics in ballistic 1D channels, which is clearly indicated by the presence of Schwinger pair production, yielding antifermion states which undergo confinement and exhibit properties which are highly suitable for applications in spintronics. In the situation described, the parameters governing the Dirac equation are highly tunable, in contrast to the situation in both high-energy physics and other emergent Dirac systems such as graphene and the topological insulators. In our case, the magnetic field plays a role analogous to the Dirac mass, the spin-orbit interaction plays the role of the speed of light, and the electric field is provided by the smooth 1D potential inside the constriction. The similarity and tunability of these energy scales allows for remarkable control over the properties of the Dirac states and enhances the versatility of the system for applications to spintronics.
 The author wishes to acknowledge O.P. Sushkov, J. Ingham, M. Veldhorst, I.S. Terekhov and A.R. Hamilton for valuable discussions.

%\bibitem{Meir}
%Y. Meir, J. Phys.: Condens. Matter {\bf 20} 164208 (2008).
%\bibitem{Thomas}
%K.J. Thomas, J.T. Nicholls, M.Y. Simmons, M. Pepper, D.R. Mace, and D.A. Ritchie, Phys. Rev. Lett. {\bf 77}, 135 (1996).
%\bibitem{Cronenwett}
%S.M. Cronenwett, \emph{et. al.}, Phys. Rev. Lett. {\bf 88}, 226805 (2002).
%\bibitem{Sarkozy}
%S. Sarkozy, \emph{et. al.}, Phys. Rev. B {\bf 79}, 161307 (2009).
%\bibitem{Ren}
%Y. Ren, W.W. Yu, S.M. Frolov, J.A. Folk, and W. Wegscheider, Phys. Rev. B {\bf 82}, 045313 (2010).
%\bibitem{Oleh}
%O. Klochan, A.P. Micolich, A.R. Hamilton, K. Trunov, D. Reuter, and A.D. Wieck, Phys. Rev. Lett. {\bf 107}, 076805 (2011).
%\bibitem{footnote}
%We take coordinates so that the 2D system lies in the $x-y$ plane, with the channel parallel to the $x$-axis.

\end{document}